\documentclass[runninghead]{llncs}

\usepackage{doi}
\usepackage{microtype}
\usepackage{enumitem}
\usepackage{float}
\usepackage{multirow}
\usepackage{booktabs}
\usepackage{siunitx}
\usepackage{graphicx}
\usepackage[T1]{fontenc}
\usepackage{cite}
\usepackage{amssymb}
\usepackage{placeins}

\begin{document}

\title{HiFiNet: Hierarchical Fault Identification in Wireless Sensor
Networks via Edge‑Based Classification and Graph Aggregation}

\titlerunning{HiFiNet}

\author{
  Nguyen Tri Nghia \inst{1} \and
  Nguyen Van Son \inst{2,*} \and
  Nguyen Thi Hanh \inst{3,*}
}

\authorrunning{Nghia Nguyen et al.}

\institute{
  Hanoi University of Science and Technology, Hanoi, Vietnam.
  \\ \email{nghia.nt215438@sis.hust.edu.vn}\and
  Phenikaa School of Computing, PHENIKAA University, Hanoi, Vietnam.
  \\ \email{son.nguyenvan@phenikaa-uni.edu.vn}\and
  Phenikaa School of Interdisciplinary Digital Technology, PHENIKAA
  University, Hanoi, Vietnam. \\ \email{hanh.nguyenthi@phenikaa-uni.edu.vn}\\
  \inst{*}Corresponding author\\
}

\maketitle

\begin{abstract}
  Wireless Sensor Networks (WSN) are the backbone of essential
  monitoring applications, but their deployment in unfavourable
  conditions increases the risk to data integrity and system
  reliability. Traditional fault detection methods often struggle to
  effectively balance accuracy and energy consumption, and they may
  not fully leverage the complex spatio-temporal correlations
  inherent in WSN data. In this paper, we introduce HiFiNet, a novel
  hierarchical fault identification framework that addresses these
  challenges through a two-stage process. Firstly, edge classifiers
  with a Long Short-Term Memory (LSTM) stacked autoencoder perform
  temporal feature extraction and output initial fault class
  prediction for individual sensor nodes. Using these results, a
  Graph Attention Network (GAT) then aggregates information from
  neighboring nodes to refine the classification by integrating the
  topology context. Our method is able to produce more accurate
  predictions by capturing both local temporal patterns and
  network-wide spatial dependencies. To validate this approach, we
  constructed synthetic WSN datasets by introducing specific,
  predefined faults into the Intel Lab Dataset and NASA's MERRA-2
  reanalysis data. Experimental results demonstrate that HiFiNet
  significantly outperforms existing methods in accuracy, F1-score,
  and precision, showcasing its robustness and effectiveness in
  identifying diverse fault types. Furthermore, the framework's
  design allows for a tunable trade-off between diagnostic
  performance and energy efficiency, making it adaptable to different
  operational requirements.

\end{abstract}

\section{Introduction}

Wireless Sensor Networks (WSN) serve as the backbone for critical
monitoring applications in fields such as healthcare, logistics, and
environmental surveillance. These networks often operate in harsh,
inaccessible environments—ranging from active volcanoes to industrial
plants with extreme conditions—where wired solutions are impractical
\cite{Yick2008,Chai2020,Gungor2009}. However, this versatility
exposes sensor nodes to resource constraints and environmental
stress, making them highly susceptible to failure. Such data faults
can lead to incorrect predictions, ultimately undermining the
reliability of the monitoring system. Ensuring data integrity is
paramount, yet traditional fault detection methods often struggle to
balance accuracy with the severe energy constraints of WSN nodes.

Early fault detection strategies relied heavily on model-based or
statistical techniques. For instance, Panda et al. utilized z-score
functions \cite{Panda2014}, while Ahmad et al. proposed Kalman
filtering for lightweight anomaly detection \cite{Ahmad2024}. While
computationally efficient, these methods typically require hand-tuned
thresholds and degrade rapidly under fluctuating environmental
conditions, leading to high false alarm rates \cite{Zhang2018}. To
address these limitations, research has pivoted toward data-driven
Machine Learning (ML) approaches, which automatically learn latent
features from historical data. Techniques such as Extremely
Randomized Trees \cite{Saeed2021} and Support Vector Machines have
demonstrated superior ability to capture complex non-linear sensor
patterns compared to static rules. However, these pure ML models
often operate as "black boxes" and can struggle to distinguish
between environmental events and sensor faults without adequate spatial context.

Furthermore, the architecture of deployment remains a critical
bottleneck. Existing frameworks generally fall into three categories:
centralized, distributed, or self-diagnosis. Centralized schemes
simplify logic but suffer from high latency and communication
bottlenecks at the sink \cite{Zhang2018}. Conversely, self-diagnosis
methods, such as the Deep Belief Network proposed by Prasad et al.
\cite{Prasad2023}, allow nodes to monitor themselves to save energy
but are prone to bias and lack network-wide perspective. Distributed
neighbor-based strategies improve accuracy by cross-referencing peers
but drastically reduce network lifespan due to the high energy cost
of constant local communication \cite{Adday2022}.

Recent hybrid attempts \cite{Shi2024} suggest fusing these
approaches, yet few frameworks successfully integrate deep temporal
feature extraction at the edge with efficient graph-based spatial
aggregation. Therefore, this paper investigates the efficacy of a
data-driven hierarchical method for WSNs fault diagnosis. The major
contributions of this paper are as follows:

\begin{itemize}
  \item Introduce a novel Iterative Graph Network (IGN) that
    integrates Graph Attention Convolution with a custom Confidence
    Modulator(CM), dynamically refining node representations through
    iterative confidence propagation.
  \item Propose HiFiNet, a hierarchical network. HiFiNet first
    employs a Long Short-Term Memory-based stacked autoencoder
    (LSTM-SAE) at the edge for temporal feature extraction and
    initial fault screening, followed by IGN for aggregating
    neighborhood information and refining fault diagnosis.
  \item Simulate real-world fault scenarios by generating synthetic
    WSN datasets that combine the Intel Lab Dataset measurements
    \cite{Intel2004} with environmental measurements drawn from
    NASA’s MERRA‑2 reanalysis data \cite{GMAO2015}.
  \item Evaluate the performance with various metrics such as
    accuracy, F1-score, and precision on the above datasets, and
    demonstrate improvement against methods in the literature.
\end{itemize}

\section{System Model and Problem Statement}

\subsection{System Model}
We consider a WSN as an undirected graph \(G=(V,E)\), where
\(V=\{v_1, v_2, \ldots, v_N\}\) is the set of \(N\) sensor nodes and
\(E \subseteq V \times V\) represents bidirectional communication
links between neighboring nodes. Each node \(v_i\) is equipped with a
sensing modality and generates a time-series measurement \(x_i(t)\)
at discrete time steps \(t=1, 2, \ldots, T\). Communication between
\(v_i\) and \(v_j\) is possible if \((v_i,v_j) \in E\).
This paper makes the following additional assumptions:
\begin{itemize}
\item Static or Slowly Varying Topology: Sensor nodes are assumed to
  be stationary or have negligible mobility during the diagnosis
  window, so that network links \(E\) remain constant or change infrequently.
\item Global Time Synchronization: Nodes maintain loosely
  synchronized clocks, ensuring measurements \(x_i(t)\) across the
  network are aligned.
\item Existence of a High-Capacity Base Node: A base node with higher
  computation and memory resources is reachable (directly or
  multi-hop) from all nodes and knows the network topology apriori.
\end{itemize}

\subsection{Problem Statement}
Given a segment of a sensor time series and a predefined set of fault
classes, the problem is to assign the most appropriate fault class
label to that time series segment.

Consider an input time series \(X_w = (x_t, x_{t+1}, \ldots,
x_{t+w-1})\) of length \(w\), where \(x_i \in \mathbb{R}\) is the
sensor reading at discrete time step \(i\). Define \(C = \{c_1, c_2,
\ldots, c_K\}\) as the set of \(K\) mutually exclusive fault classes
as detailed in the fault taxonomy in Section~\ref{subsec:types} and
the normal (fault-free) class.

The object is to learn a classification function (or hypothesis)
\(h\) that maps the input feature vector \(\mathbf{x}\) (derived from
\(X_w\)) to a predicted fault class label \(\hat{y}\ \in C\).

\subsection{Fault Taxonomy}
\label{subsec:types}
We adopt the characteristic-based fault taxonomy defined in WSN
literatures \cite{Saeed2021,Shi2024,Hasan2024}, focusing on five
types: Hardover (constant bias), Drift (linear deviation), Spike
(transient impulse), Erratic (increased variance), and Stuck-at
(fixed value). Synthetic faults were injected into the dataset
following the mathematical models detailed in \cite{Saeed2021}.

\section{Proposed Algorithm}
\label{sec:method}

Our goal is to detect and label faults in wireless sensor networks
(WSNs) from each node's time-series data and its spatial context. The
expected output is a fault label (or "normal") for every data segment.

\begin{figure}
  \centering
  \includegraphics[width=0.8\linewidth]{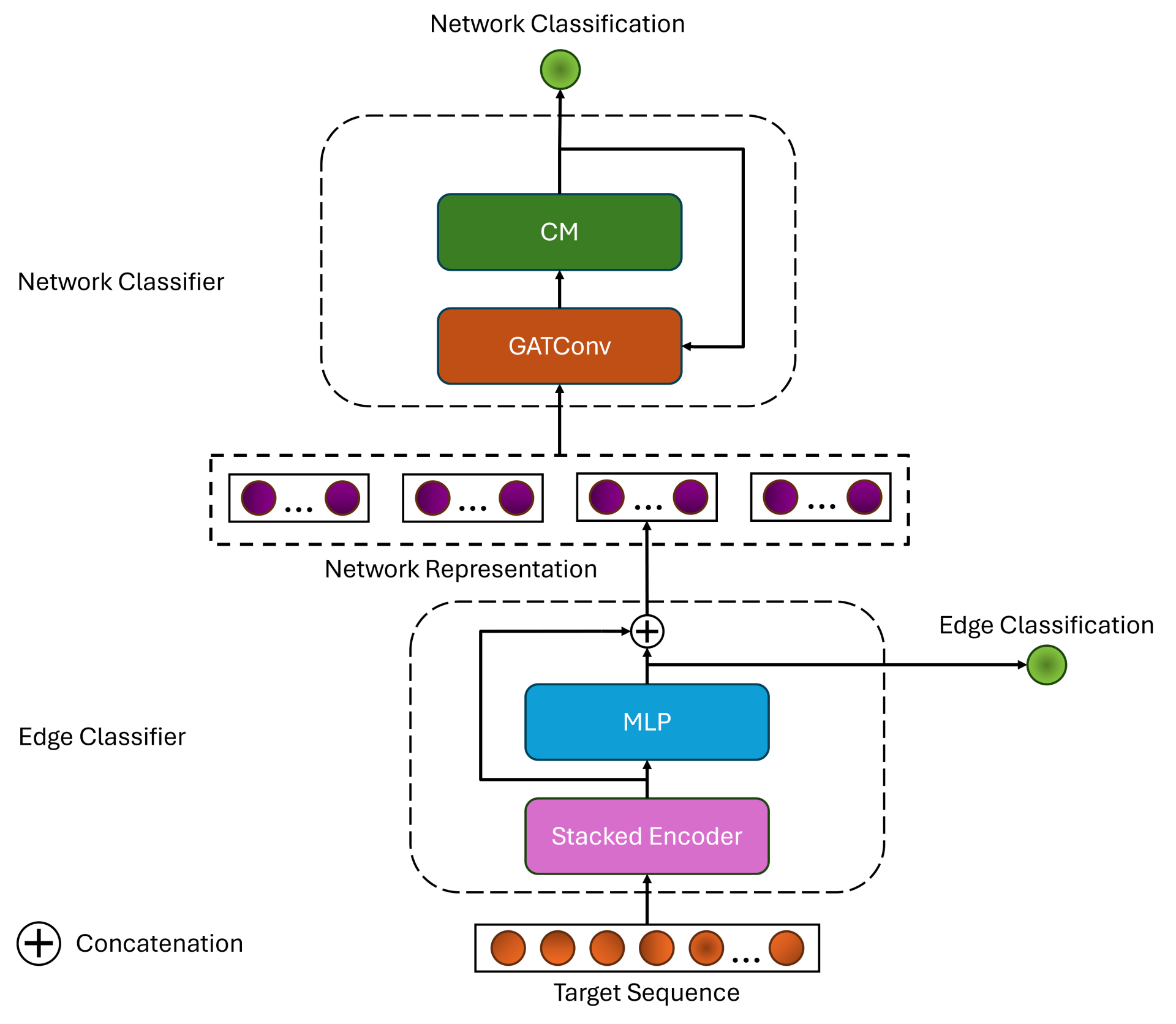}
  \caption{Proposed HiFiNet inference pipeline, illustrating the Edge
    Classifier processing a target node's temperature sequence and the
  Network Classifier integrating edge outputs with contextual network data.}
  \label{fig:hifinet}
\end{figure}

We propose HiFiNet, a Hierarchical Fault Identification Network, for
the classification of sensor faults in WSNs. HiFiNet jointly
leverages per-node temporal features and network‐level interactions
to outperform single-level baselines. Moving beyond conventional
approaches, our key innovations lie in the novel integration of a
confidence-guided Graph Attention Network. This architecture
establishes a new paradigm for fault diagnosis by dynamically
learning to trust and weigh information from individual sensors and
the broader network context. The overall inference pipeline is
depicted in Figure~\ref{fig:hifinet}.

\subsection{Edge Classifier}
The first stage of HiFiNet is the Edge Classifier. This component's
primary role is to analyze the time-series data from an individual
target sensor node. This analysis extracts discriminative features
and performs an initial fault classification. The Edge Classifier is
pre-trained using a standard stacked LSTM autoencoder approach to
minimize reconstruction loss, followed by supervised fine-tuning. The
edge logits and the learned embedding are then forwarded to the
network layer to serve as the initial node states, denoted as
\(\mathcal{H}^{(0)}\).

\subsection{Iterative Graph Network}
\begin{figure}
  \centering
  \includegraphics[width=0.6\linewidth]{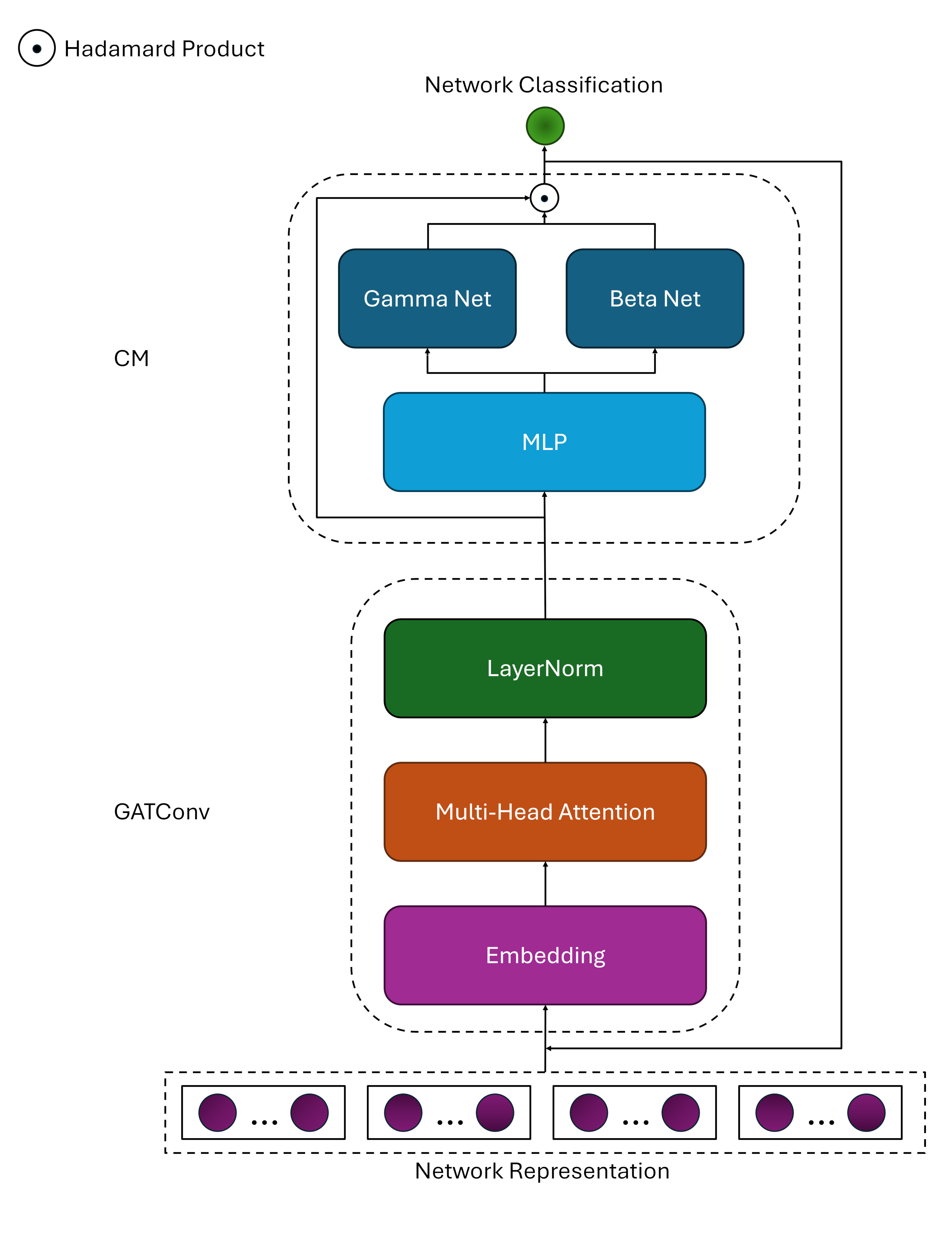}
  \caption{Illustration of the Iterative Graph Network architecture,
    showing the iterative process of feature modulation, graph
  attention convolution, and confidence update.}
  \label{fig:ign}
\end{figure}

The second stage of HiFiNet is the IGN, which refines the initial
fault assessments or embeddings from the Edge Classifier by
incorporating network-wide spatial context and internode
dependencies. The core of the Network Classifier is an IGN.
Figure~\ref{fig:ign} depicts the information flow in an IGN.

The IGN is designed to dynamically refine node representations by
iteratively propagating and aggregating information across the graph,
modulated by an evolving confidence measure. Let \(\mathcal{H}^{(0)}
\in \mathbb{R}^{N \times d_0}\) be the vector representation of a
node with \(N\) is the number of nodes and \(d_0\) is the dimension
of the edge output. The IGN performs \(K\) iterations.

In each iteration \(k = 0, \dots, K\):
\begin{itemize}
  \item Feature Modulation (for \(k > 0\)):
    The initial node embeddings \(\mathbf{H}^{(0)}\) are modulated
    based on a confidence score vector \(\mathbb{c}^{(k-1)} \in
    \mathbb{R}^{N \times 1}\), derived from the \((k-1)\)-th
    iteration's GAT output. A Confidence Modulator function,
    \(\mathcal{M}_{\text{c}}\), is employed using Feature-wise Linear
    Modulation (FiLM):
    \begin{equation}
      \mathcal{H}'^{(k)} = \mathcal{M}_{\text{c}}(\mathcal{H}^{(0)},
      \mathbb{c}^{(k-1)}) = g_{\gamma}(c^{(k-1)}) \odot
      \mathcal{H}^{(0)} + g_{\beta}(c^{(k-1)})
    \end{equation}
    where \(g_{\gamma}(\cdot)\) and \(g_{\beta}(\cdot)\) are
    learnable functions that generate scaling and shifting parameters
    from the confidence scores, and \(\odot\) denotes element-wise
    multiplication. For the first iteration \(k=0\), no modulation is
    applied, so \(\mathcal{H}'^{(0)} = \mathcal{H}^{(0)}\). The
    Confidence Modulator allows the network to adaptively emphasize
    or de-emphasize features based on the current confidence in their
    classification.

  \item Graph Attention Convolution (GAT) Block:
    The modulated features \({\mathbb{H}'}^{(k)}\) are then processed
    by a GAT block, denoted as \(\mathcal{G}\). This block typically
    consists of one or more Graph Attention Convolution layers, which
    enable nodes to selectively attend to their neighbors' features
    when updating their own representations:
    \begin{equation}
      \mathcal{H}_{\text{G}}^{(k)} = \mathcal{G}(\mathcal{H}'^{(k)}, A)
    \end{equation}
    where \(A\) is the adjacency matrix representing the graph
    structure. Each GAT layer computes attention coefficients for
    neighboring nodes, aggregates their features weighted by these
    coefficients, and applies a linear transformation. Layer
    normalization and activation functions are often applied between
    GAT layers. Dropout are also used for regularization. The output
    \(\mathcal{H}_{\text{G}}^{(k)} \in \mathbb{R}^{N \times
    d_{\text{G}}}\) contains node representations that have
    incorporated information from their local graph neighborhood.

  \item Temporary Classification and Confidence Update (if \(k < K-1\)):
    If it is not the final iteration, the output
    \(\mathcal{H}_{\text{G}}^{(k)}\) is passed through a temporary
    classifier, \(f_{\text{temp}}\), to obtain intermediate class
    logits \(z^{(k)}\) for each node:
    \begin{equation}
      z^{(k)} = f_{\text{temp}}(\mathcal{H}_{\text{G}}^{(k)})
    \end{equation}
    These logits are converted to probabilities \(P^{(k)}\) using the
    softmax function:
    \begin{equation}
      P^{(k)} = \text{softmax}(\mathbf{z}^{(k)})
    \end{equation}
    The confidence score \(c_i^{(k)}\) for each node \(i\) is then
    derived from these probabilities, typically as the maximum
    probability value:
    \begin{equation}
      c_i^{(k)} = \max(P_i^{(k)})
    \end{equation}
    This results in a confidence vector \(\mathbf{c}^{(k)}\), which
    is then used in the Feature Modulation step of the next iteration
    (\(k+1\)). This iterative process allows the model to
    progressively refine its understanding by focusing subsequent GAT
    operations based on the certainty of intermediate predictions.
\end{itemize}

After \(K\) iterations, the final GAT output
\(\mathcal{H}_{\text{G}}^{(K_{iter}-1)}\) represents the
spatially-refined node embeddings. This and the original
representation are then passed to a classification head to output the
network classification. In inference, only the encoder and the IGN
forward pass are needed; the reconstruction branch is omitted.

\section{Experiments and Result}

In this section, we evaluated our proposed method against other
methods in the literature. From previous research about fault
diagnosis in WSNs, we selected  Deep Belief Network
\cite{Prasad2023}, LSTM-AE \cite{Khan2024}, and Support Vector
Machine(SVM). We utilized two datasets for evaluation: NASA MERRA
\cite{GMAO2015}  (reanalysis temperature data from Northern Vietnam)
and Intel Lab Dataset \cite{Intel2004} (WSN from an Intel lab).

\subsection{Fault Injection}
A key challenge in developing fault detection systems in WSNs is the
scarcity of real-world fault data, making it difficult to test a
model's robustness. Therefore, to evaluate the performance of HiFiNet
in realistic environment, synthetic faults were injected into the
pre-processed normal temperature data.

Multiple datasets were generated with varying overall percentages of
faulty instances, specifically 5\%, 10\%, 15\%, and 20\% of the total
data points being faulty. Analysis of real-world data has shown
\(<20\%\) readings faults in WSNs. Within each dataset, the total
fault rate was divided equally among the five predefined fault types.
In total, four datasets are created from each original dataset
corresponding to each fault rate.

\subsection{Evaluation Result}
\begin{figure}
  \centering
  \includegraphics[width=\linewidth]{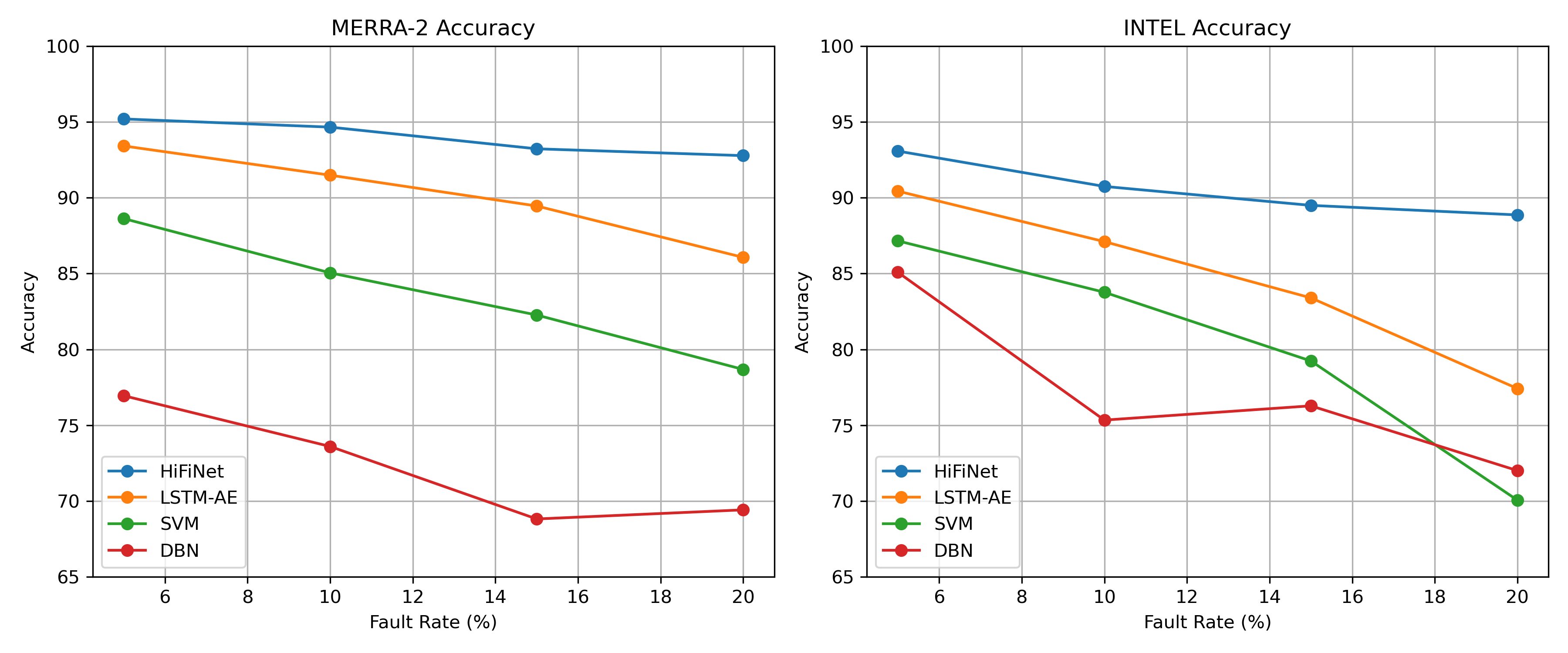}
  \caption{Accuracy versus Fault Rate comparison between HiFiNet and
  other methods.}
  \label{fig:accuracy}
\end{figure}

We now verify the two claims in Section~\ref{sec:method}: (i)
edge-to-graph synergy, (ii) energy-aware tunability. The overall
classification accuracy, shown in Figure~\ref{fig:accuracy},
demonstrates a substantial uplift from HiFiNet. HiFiNet maintains a
\(2\sim6\%\) accuracy advantage over the second-best model LSTM-AE.
SVM and DBN lag behind with sub \(90\%\) and sub \(80\%\) accuracies,
respectively, across datasets. More importantly, the accuracy of
HiFiNet does not degrade significantly as the fault rate increases
compared to other methods. This indicates that information from
neighbors collected in HiFiNet has a major impact on stabilizing the
performance when noise becomes more prevalent

\begin{figure}
  \centering
  \includegraphics[width=0.7\linewidth]{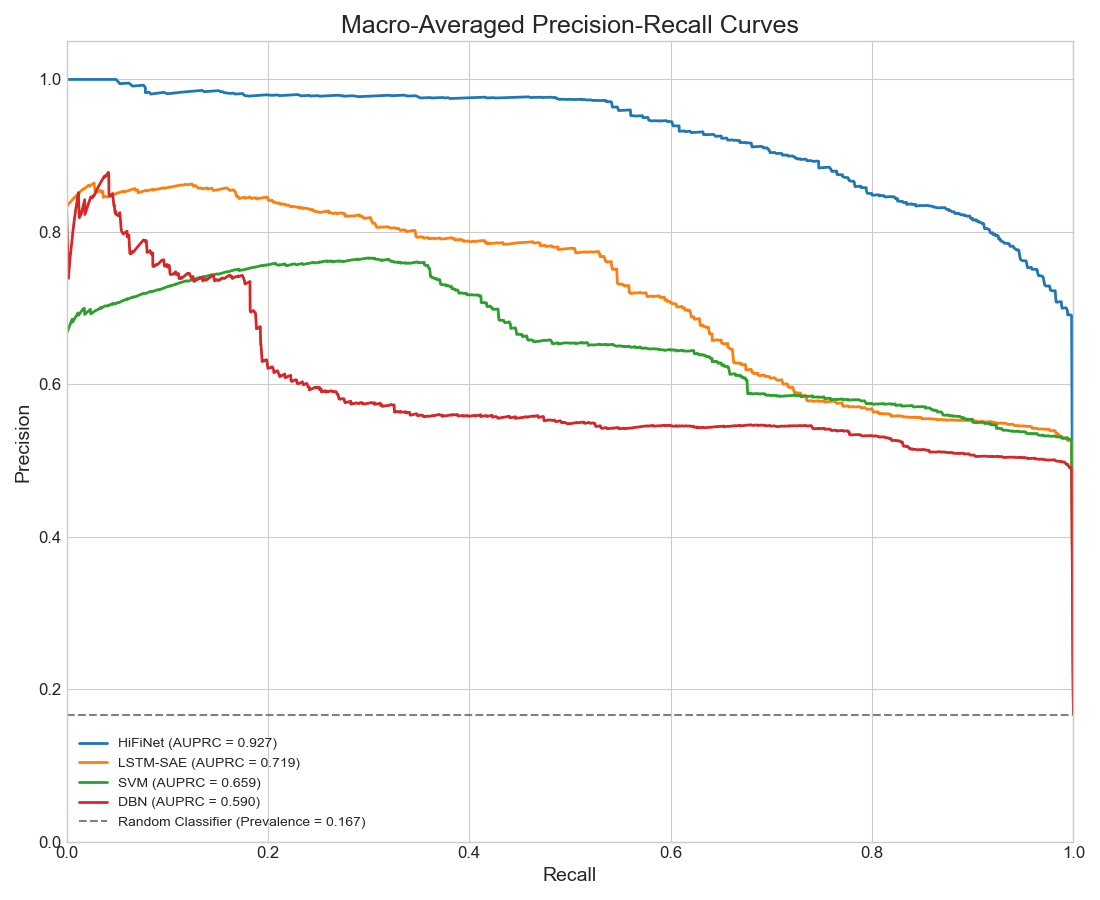}
  \caption{Precision-Recall curves comparison between HiFiNet and
  other methods on Intel 20\% dataset.}
  \label{fig:pr_curve}
\end{figure}

To further analyze each method's discriminative performance,
Figure~\ref{fig:pr_curve} plots the Precision-Recall curve at the
highest stress level (20\% fault rate). The results clearly indicate
that the HiFiNet model substantially outperforms the other methods,
achieving an AUPRC of 0.927. Its curve dominates the others,
maintaining high precision (above 0.95) for recall values up to
approximately 0.8. The next best performing model is the LSTM-SAE,
with an AUPRC of 0.719, followed by the SVM at 0.659 and the DBN at
0.590. Overall, the HiFiNet architecture demonstrates a superior
balance of precision and recall for this classification task.

\begin{figure}
  \centering
  \includegraphics[width=\linewidth]{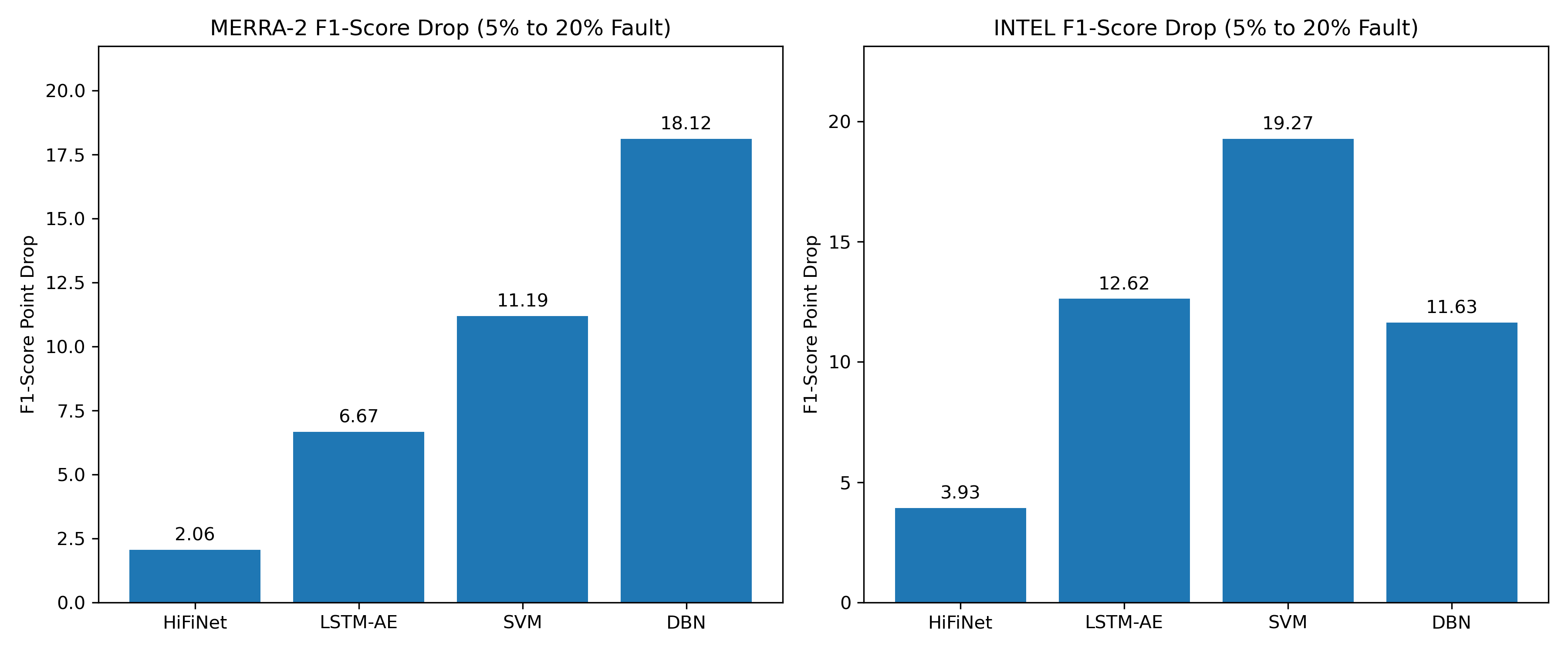}
  \caption{F1-Score drop comparison between HiFiNet and other methods.}
  \label{fig:f1_drop}
\end{figure}

Next, Figure~\ref{fig:f1_drop} provides a direct, quantitative
measure of model stability. It visualizes the absolute drop in
F1-Score as the fault rate increases from 5\% to 20\%. A smaller bar
indicates greater robustness against worsening data quality. HiFiNet
clearly stands out with the smallest performance drop on both
datasets (2.06 points for MERRA-2 and 3.93 for INTEL). In contrast,
models like SVM and LSTM-AE experience much larger drops, exceeding
11 and 12 points, respectively. This bar chart offers a stark visual
confirmation of HiFiNet's superior stability and reliability under
varying conditions.

\begin{figure}
  \centering
  \includegraphics[width=0.8\linewidth]{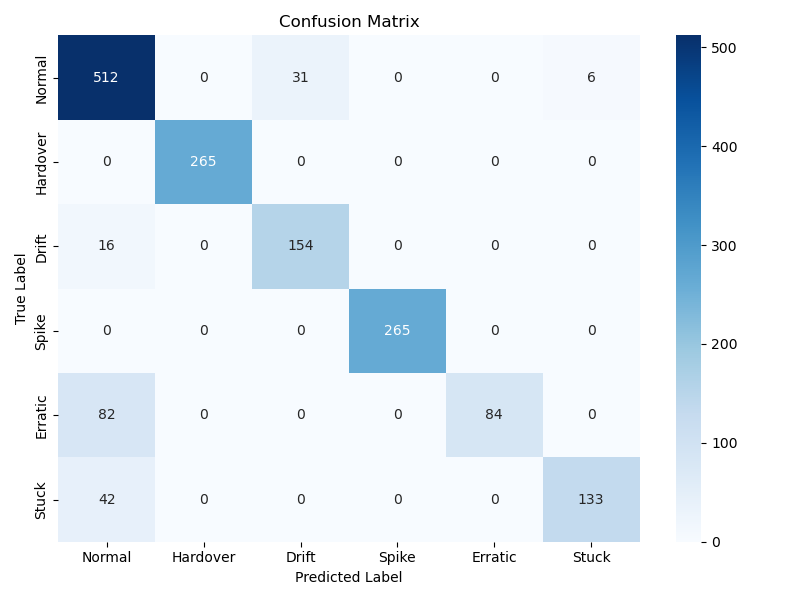}
  \caption{Confusion Matrix of HiFiNet on Intel 20\% dataset.}
  \label{fig:confusion_matrix}
\end{figure}

To further analyze HiFiNet's performance on a granular level,
Figure~\ref{fig:confusion_matrix} provides the confusion matrix of
the model. HiFiNet demonstrates high accuracy for several categories,
correctly identifying the majority of instances for Normal, Hardover,
Drift, and Spike classes. However, performance on Stuck-at fault
remains less than desirable, with 42 misclassifications out of 175
samples. The model also has struggles with identifying Erratic
faults, with almost half of the instances in the class being misclassifications.

\subsection{Ablation Study: Accuracy Tradeoff versus Energy
Efficiency of HiFiNet}
We investigated the trade-off between the accuracy gained from
spatial aggregation and the associated energy costs. Energy
consumption was quantified using the standard First Order Radio
Model~\cite{Heinzelman2000} with \(\epsilon_{elec} = 50\) nJ/bit and
\(\epsilon_{fs} = 10$ pJ/bit/m$^2\). We introduced a tunable
parameter, \textit{Time Delay} (\(t\)), which controls the frequency
of the Network Classifier's execution. A delay of \(t=k\) implies
that spatial aggregation is performed only once every \(+1\) time windows.

\begin{figure}
  \centering
  \includegraphics[width=\linewidth]{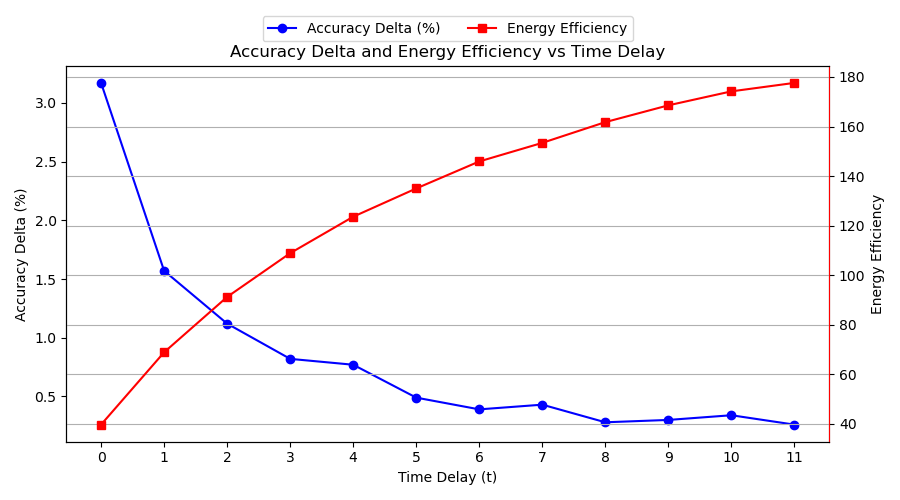}
  \caption{Accuracy Delta and Energy Efficiency versus Time Delay.
    Accuracy Delta represents the performance gain from using the
    Network Classifier compared to the Edge Classifier alone. Energy
    Efficiency is inversely proportional to the energy consumed by the
  communication process.}
  \label{fig:accuracy_tradeoff}
\end{figure}

Figure~\ref{fig:accuracy_tradeoff} illustrates the results. At
\(t=0\), HiFiNet achieves the maximum diagnostic benefit \(+3.1\%\)
accuracy over the edge-only baseline but results in the lowest energy
efficiency. As \(t\) increases, energy efficiency rises drastically,
improving by a factor of roughly 10 at \(t=8\), while the accuracy
gain gradually diminishes. This demonstrates HiFiNet's adaptability:
operators can select a low \(t\) for critical, high-precision tasks,
or a high \(t\) (\(t \ge 4\)) to maximize network longevity with only
a marginal drop in fault detection performance.

\FloatBarrier

\section{Conclusion}

In this paper, we have presented HiFiNet, a hierarchical fault
identification network for Wireless Sensor Networks that
synergistically combines edge-based classification with graph
aggregation. By first analyzing temporal data at the individual node
level and then refining the diagnosis through a graph attention
mechanism that incorporates spatial context from neighboring nodes,
HiFiNet achieves a new level of accuracy and robustness in fault
detection. Our experiments have demonstrated the superiority of
HiFiNet over methods such as DBN, LSTM-AE, and SVM. The results show
a significant improvement in key performance metrics, including
accuracy, F1-score, and precision, across various fault types and
rates. Notably, HiFiNet exhibits greater stability as noise and data
quality issues increase, a crucial advantage in real-world WSN
deployments. A tunable mechanism unique to HiFiNet was also explored,
which demonstrated the model's ability to balance accuracy and power
consumption.

\section*{Acknowledgement}
This research is funded by Vietnam National Foundation for Science
and Technology Development (NAFOSTED) under grant number 102.01-2025.69.

\bibliography{references}

\end{document}